\documentclass[twocolumn,prb,aps,amsfonts,amsmath,graphics]{revtex4}
\usepackage{graphics}
\usepackage{graphicx}
\usepackage{amssymb}

\draft
\begin{document}
\draft
\author{M.J.Everitt, P.Stiffell, T.D.Clark\thanks{
e-mail address: t.d.clark@sussex.ac.uk},A.Vourdas$^{\dag}$
\linebreak J.F.Ralph\thanks{
Department of Electrical Engineering and Electronics, Liverpool University,
Brownlow Hill, Liverpool L69 3GJ, U.K.}, H.Prance and R.J.Prance.}
\address{Quantum Circuits Research Group, School of Engineering, University of Sussex,
\\
Brighton, Sussex BN1 9QT, U.K.}
\title{A Fully Quantum Mechanical Model of a SQUID Ring Coupled\\
to an Electromagnetic Field}
\pacs{85.25.Dq  03.65.-w  42.50.Dv}

\begin{abstract}
A quantum system comprising of a monochromatic electromagnetic field coupled
to a SQUID ring with sinusoidal non-linearity, is studied. A magnetostatic
flux $\Phi_x$ is also threading the SQUID ring, and is used to control the
coupling between the two systems. It is shown that for special values of $
\Phi_x$ the system is strongly coupled. The time evolution of the system is
studied. It is shown that exchange of energy takes place between the two
modes and that the system becomes entangled. A second quasi-classical model
that treats the electromagnetic field classically is also studied. A
comparison between the fully quantum mechanical model with the
electromagnetic field initially in a coherent state and the quasi-classical
model, is made.
\end{abstract}
\maketitle


\section{Introduction}
With the SQUID ring (here taken to be a thick superconducting ring enclosing
a single Josephson weak link device) regarded as having potential for future
quantum technologies~\cite{1.,2.,schon1999,A}, it is clearly of interest to
consider its interaction with an external quantum mechanical electromagnetic
(em) field. This interest has certainly been promoted by the recent
experimental work on the creation of quantum mechanical superposition states
of Josephson systems~\cite{R,S,Na,N}, with particular emphasis on the
existence of such states in SQUID rings~\cite{L,M}. As with these latter
experiments, in order to investigate these states we consider a
monochromatic em field with frequency $\omega _{e}$ (typically in the 0.1 to
1 THz region), coupled to a SQUID ring oscillator with frequency $\omega
_{s} $. In addition a magnetostatic flux $\Phi _{x}$ is also applied to the
ring, as depicted in figure~1. Since the primary purpose of the work
reported here is to study the full quantum mechanics of this coupled system,
we make the assumption that the operating temperature $\left( T\right) $ is
such that $\hbar \omega _{e}\gg k_{B}T$, $\hbar \omega _{s}\gg k_{B}T$ so
that both the ring and field modes behave quantum mechanically.

As we shall show, in this fully quantum mechanical description the quantum
states of the em field mode plus SQUID ring couple together strongly only
under certain circumstances, specifically around particular values of the
magnetostatic bias flux $\Phi _{x}$. In this case, using the bias flux as a
means to control the coupling, we have been able to reveal a whole range of
interesting quantum phenomena.

\begin{figure}
\protect{  \begin{center}
    \resizebox*{0.48\textwidth}{!}{\includegraphics{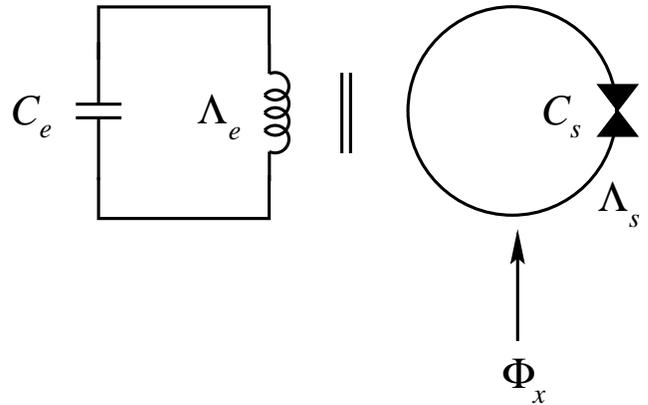}}
  \end{center} 
  \caption{
A SQUID ring coupled inductively to a mode of an
electromagnetic field.
 \label{fig:1}
 }}
\end{figure}

In previous work~\cite{3.,3a.,3b.} we dealt with the semi-classical problem
of a monochromatic microwave field coupled to a SQUID ring containing a
small capacitance weak link. In this paper we extend our theoretical
description and treat both the ring and the field fully within a quantum
mechanical framework. We demonstrate that the numerical results derived from
this quantum model, in which the em field is initially in a coherent state,
compare very well with those obtained using a semi-classical, Floquet theory
of a SQUID ring coupled to the field. In this there are obvious analogies to
quantum optical interactions in few level atoms which apply to both pair
condensate and single electron systems~\cite
{schon1992,schon2000,kastner1992,spiller1992,grabert1992}. In addition, we
note that SQUID rings have a strong sinusoidal non-linearity and it is the
strength of this non-linearity, together with its periodic nature, that
leads to the quite novel phenomena studied in this paper. This should be
compared and contrasted with the large body of work on non-linear quantum
systems in the context of quantum optics~\cite
{wod85,yurke86,campos89,fearn89,singer90,vourdas92} where the non-linearity
is usually a weak polynomial non-linearity.

\section{A SQUID ring coupled to non-classical em field}

The Hamiltonian $H_{t}$ for our coupled system can be written as a sum of
the energies for the field and the ring, together with an additional term
for the interaction energy, i.e. 
\begin{equation}
H_{t}=H_{e}+H_{s}-H_{{\rm Int}}.  \label{eq:total}
\end{equation}
where $H_{e}$ and $H_{s}$ are, respectively, the Hamiltonians for the field
and the ring and $H_{{\rm Int}}$ is the interaction energy.

We can write the Hamiltonian for the SQUID ring (weak link capacitance $
C_{s} $ and ring inductance $\Lambda _{s}$), in the usual form~\cite
{spiller1992} 
\begin{equation}
H_{s}=\frac{Q_{s}^{2}}{2C_{s}}+\frac{\left( \Phi _{s}-\Phi _{x}\right) ^{2}}{
2\Lambda _{s}}-\hbar \nu \cos \left( 2\pi \frac{\Phi _{s}}{\Phi _{0}}\right)
.  \label{eq:HamS}
\end{equation}
where $\Phi _{s}$, the magnetic flux threading the ring, and $Q_{s}$, the
total charge across the weak link, are the conjugate variables for the
system (with the imposed commutation relation $\left[ \Phi _{s},Q_{s}\right]
=i\hbar $), $\Phi _{x}$ is the static (or quasi-static) external flux
applied to the ring, $\hbar \nu /2$ is the matrix element for pair
tunnelling through the weak link (critical current $I_{c}=2e\nu $) and $\Phi
_{0}=h/2e$. We note that with a characteristic frequency $\omega _{s}=\left(
1/\sqrt{C_{s}\Lambda _{s}}\right) $ for the SQUID ring, there is a
renormalized frequency $\Omega _{s}=\omega _{s}+4\hbar ^{2}\pi ^{2}\nu \Phi
_{0}^{-2}C_{s}^{-1}\omega _{s}^{-1}$ related to the $\Phi _{s}^{2}$ term in
a Taylor expansion of the cosine in (\ref{eq:HamS}). Throughout the paper we
use $C_{s}=1\times 10^{-16}$F, $\Lambda _{s}=3\times 10^{-10}$H and $\hbar
\nu =0.07\Phi _{0}^{2}/\Lambda _{s}$ as typical circuit parameters for a
SQUID ring in the quantum regime.

The em field can be modelled in terms of a cavity mode using an equivalent
circuit comprising a capacitance $C_{e}$ in parallel with an inductance $
\Lambda _{e}$, with a (parallel) resistance on resonance to define its
quality factor. If we assume this resistance to be infinite we obtain a
Hamiltonian for the field in terms of the equivalent circuit flux and charge
operators 
\begin{equation}
H_{e}=\frac{Q_{e}^{2}}{2C_{e}}+\frac{\Phi _{e}^{2}}{2\Lambda _{e}}.
\label{eq:HamM}
\end{equation}
where $\Phi _{e}$ and $Q_{e}$ are, respectively, the magnetic flux and
electric charge associated with the cavity. The field frequency is $\omega
_{e}=1/\sqrt{C_{e}\Lambda _{e}}$. For the purposes of simplicity we use $
C_{e}=C_{s}$ throughout this paper and specify the frequency $\omega _{e}$
in each example. We denote as $|n\rangle $ the eigenstates of $H_{e}.$ In
our numerical work we use a truncated basis with $n=0,...,N$, where $N$ is
taken to be much greater than the average number of photons in the system.

The em cavity mode and the SQUID ring are coupled together inductively with
a coupling energy given by 
\begin{equation}
H_{{\rm Int}}=\frac{\mu }{\Lambda _{s}}\left( \Phi _{s}-\Phi _{x}\right)
\Phi _{e}  \label{eq:HamC}
\end{equation}
where $\mu $ is a coupling parameter linking the em field to the SQUID ring.

We note that by introducing a unitary translation operator ${\Bbb {T}}=\exp
\left( -i\Phi _{x}Q_{s}/\hbar \right) $ we can write the Hamiltonian for the
ring as 
\begin{equation}
{H}_{s}^{\prime }={\Bbb {T}}^{\dagger }H_{s}{\Bbb {T}}=\frac{Q_{s}^{2}}{
2C_{s}}+\frac{\Phi _{s}^{2}}{2\Lambda _{s}}-\hbar \nu \cos \left( 2\pi \frac{
\Phi _{s}+\Phi _{x}}{\Phi _{0}}\right)  \label{eq:HamST}
\end{equation}
\begin{figure}
\protect{  \begin{center}
    \resizebox*{0.48\textwidth}{!}{\includegraphics{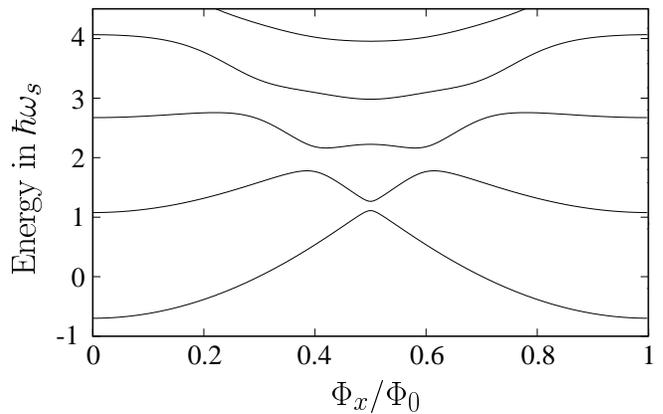}}
  \end{center} 
  \caption{
 Energy eigenvalues versus $\varphi _{x}=\Phi _{x}/\Phi _{0}$
for an isolated SQUID ring.
 \label{fig:2}
 }}
\end{figure}
We also note that, invoking this unitary transformation, the interaction
energy becomes $H_{{\rm Int}}^{\prime }=\frac{\mu } {\Lambda _{s}}\Phi
_{s}\Phi _{e}$ whilst the em field Hamiltonian remains unaffected. We denote
as $|\sigma \rangle $ the (flux-dependent) eigenstates of ${H^{\prime }}_{s}$
. Again in our numerical work we use a truncated basis with $\sigma
=0,...,\Sigma $, where $\Sigma $ is taken to be much greater than the
average energy level in which the SQUID operates. The first few eigenvalues $
\left( \sigma =0,.....4\right) $ of $H_{s}^{\prime }$ as functions of $\Phi
_{x}/\Phi _{0}\left( =\varphi _{x}\right) $ are shown in figure~2. As can be
seen, although all the eigenvalues are $\Phi _{0}$ -periodic in $\Phi _{x}$,
each displays a distinctive functional form in $\Phi _{x}$. It will become
apparent in the following discussion that these functional forms take on
great importance in determining the behaviour of the coupled system at
particular points in external bias flux.

In describing the coupled system, we now introduce the dimensionless
operators $x_{e}=\sqrt{C_{e}\omega _{e}/\hbar }\Phi _{e}$, $p_{e}=\sqrt{
1/C_{e}\hbar \omega _{e}}Q_{e}$, $x_{s}=\sqrt{C_{s}\omega _{s}/\hbar }\Phi
_{s}$ and $p_{s}=\sqrt{1/C_{s}\hbar \omega _{s}}Q_{s}$, together with the
lowering and raising operators $a_{s}=\frac{1}{\sqrt{2}}\left(
x_{s}+ip_{s}\right) $, $a_{s}^{\dag }=\frac{1}{\sqrt{2}}\left(
x_{s}-ip_{s}\right) $ for the ring and $a_{e}=\frac{1}{\sqrt{2}}\left(
x_{e}+ip_{e}\right) $, $a_{e}^{\dag }=\frac{1}{\sqrt{2}}\left(
x_{e}-ip_{e}\right) $ for the field. In terms of these operators the
Hamiltonian $H_{t}^{\prime }={\Bbb {T}}^{\dagger }H_{t}{\Bbb {T}}$ for the
coupled system (see (\ref{eq:total})) can be rewritten in the form 
\begin{eqnarray}
H_{t}^{\prime } &=&\hbar \omega _{e}\left( a_{e}^{\dag }a_{e}+\frac{1}{2}
\right) +\hbar \omega _{s}\left( a_{s}^{\dag }a_{s}+\frac{1}{2}\right) - 
\nonumber \\
&&\hbar \nu \cos \left( \frac{2\pi }{\Phi _{0}}\sqrt{\frac{\hbar }{
C_{s}\omega _{s}}}x_{s}+2\pi \varphi _{x}\right) -  \nonumber \\
&&\frac{\mu }{\Lambda _{s}}\sqrt{\frac{\hbar ^{2}}{4C_{s}C_{e}\omega
_{s}\omega _{e}}}\left( a_{s}^{\dag }+a_{s}\right) \left( a_{e}^{\dag
}+a_{e}\right)  \label{eq:HamMSCnorm1}
\end{eqnarray}

\begin{figure}
\protect{  \begin{center}
    \resizebox*{0.48\textwidth}{!}{\includegraphics{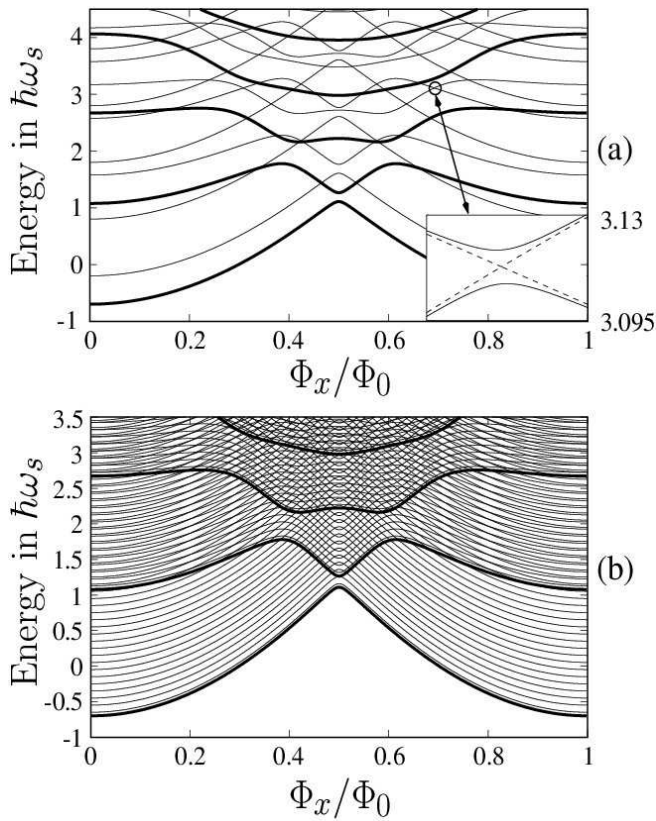}}
  \end{center} 
  \caption{
 (a) Energy eigenvalues versus $\varphi _{x}$ of the SQUID
ring Hamiltonian $H_{s}$ (thick lines) and the ring-field total Hamiltonian $
H_{t}$ (thin lines) with $\omega _{e}=\omega _{s}$. The coupling constant $
\mu =1/100$. The inset shows an example (arrowed) of the lifting of the
degeneracy of the ring-field levels when $\mu \neq 0$. (b) as in figure(a)
but with $\omega _{e}=\frac{1}{10}\omega _{s}$.
 \label{fig:3}
 }}
\end{figure}

As an illustrative example we show in figures~3 the computed, $\varphi _{x}$
-dependent eigenvalues of $H_{t}^{\prime }$ for (a) $\omega _{e}=\omega _{s}$
(with truncations $N=\Sigma =5$) and (b) $\omega _{e}=\frac{1}{10}\omega
_{s} $ (with truncations $N=50$ and $\Sigma =5$ ). In these figures the
scaling is too small to reveal the lifting of the degeneracy at the crossing
points by the nonzero coupling term $\left( \mu \neq 0\right) $. Again to
illustrate, we show in the inset of figure~3(a), but at much higher
resolution, one such computed crossing point. Here the splitting of the
crossing energies is quite apparent, these being intimately connected to the
functional form of the original SQUID ring eigenenergies. That such crossing
points have been reported in experimental studies of Josephson weak link
circuits, particularly SQUID rings, with concomitant superpositions of
macroscopic states (Schrodinger cats), is further evidence for the
underlying quantum mechanical nature of these systems~\cite{S,N,L,M}. It is
therefore timely to develop a full quantum treatment of SQUID ring-em field
systems, which is the purpose of the paper.

In the above we have studied the eigenproblem ${{H_{t}^{\prime }}} \left|
\xi _{n}\right\rangle =\Xi _{n}\left| \xi _{n}\right\rangle $ using a
truncated basis. We use now these results to compute the evolution operator
as 
\begin{equation}
U\left( t\right) =\sum_{n}\left| \xi _{n}\right\rangle \exp \left( -\frac{
i\Xi _{n}t}{\hbar }\right) \left\langle \xi _{n}\right|  \label{eq:evolution}
\end{equation}
Assuming that the system at $t=0$ is described by the density matrix $\rho
(0)$, we have calculated the density matrix $\rho (t)=U(t)\rho (0)U^{\dagger
}(t)$ at a later time $t$ and the reduced density matrices $\rho _{e}={\rm 
Tr }_{{s}}\left( \rho \right) $, $\rho _{s}={\rm Tr}_{{e}}\left( \rho
\right) $. As a measure of the accuracy of the truncation approximation we
have also calculated the traces of all the density matrices that we use. In
the limit of infinite order density matrices the trace is equal to $1$,
while for truncated density matrices it should be very close to $1$. In all
our results the trace was greater than $0.99$. Another test we performed was
to increase the cutoff point from which we were able to ascertain that our
truncation had negligible effect.

\begin{figure}
\protect{  \begin{center}
    \resizebox*{0.48\textwidth}{!}{\includegraphics{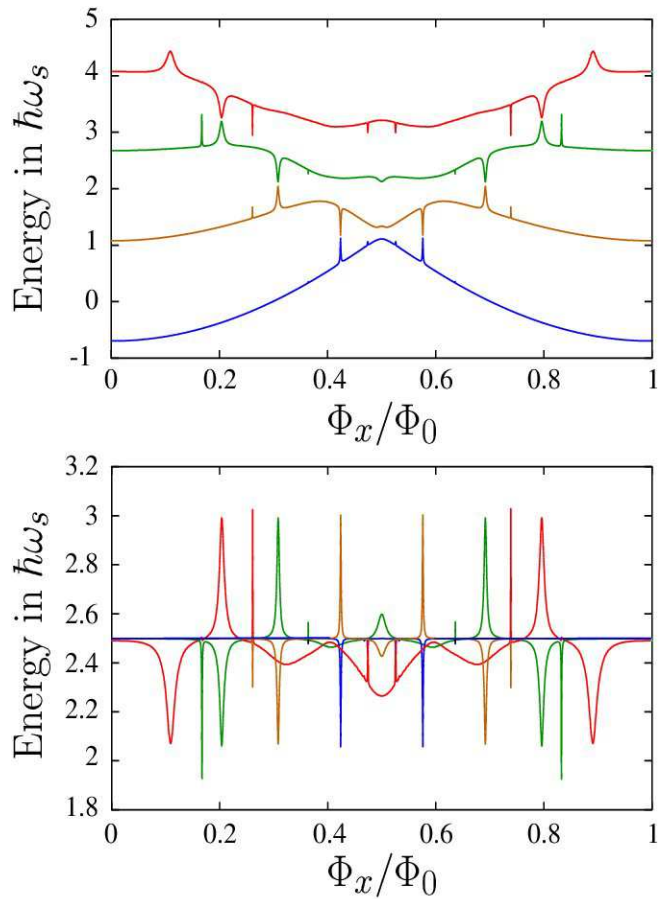}}
  \end{center} 
  \caption{
 The time averaged energy levels versus $\varphi _{x}$ for
(a) the ring ($\left\langle H_{s}\right\rangle $)and (b) the field ($
\left\langle H_{e}\right\rangle $). The coupling constant $\mu =1/100$ and $
\omega _{e}=\omega _{s}$. At $t=0$ the electromagnetic field is assumed to
be in the coherent state $|\alpha =i\sqrt{2}\rangle $ and the ring in the
energy eigenstates: $\sigma =0$ (blue), $\sigma =1$ (brown), $\sigma =2$
(green) and $\sigma =3$ (red).
 \label{fig:4}
 }}
\end{figure}

\begin{figure}
\protect{  \begin{center}
    \resizebox*{0.48\textwidth}{!}{\includegraphics{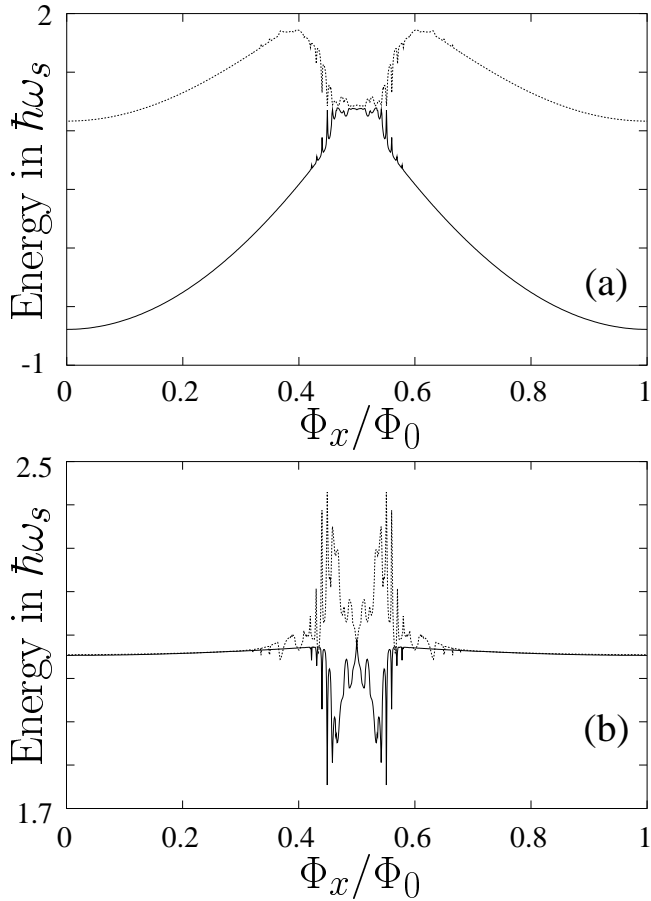}}
  \end{center} 
  \caption{
 The time averaged energy levels versus $\varphi _{x}$ for
(a) the ring ($\left\langle H_{s}\right\rangle $)and (b) the field ($
\left\langle H_{e}\right\rangle $). The coupling constant $\mu =1/100$ and $
10\omega _{e}=\omega _{s}$. At $t=0$ the electromagnetic field is assumed to
be in the coherent state $|\alpha =i10\sqrt{2}\rangle $ and the ring in the
energy eigenstates: $\sigma =0$ (solid), $\sigma =1$ (dashed).
 \label{fig:5}
 }}
\end{figure}

We define the time averaged energy expectation values $H{{^{\prime }}}_{{s}}$
and $H_{{e}}$ as $\left( i=s,e\right) $ 
\begin{equation}
\langle \langle H_{i}\rangle \rangle =\lim_{T\rightarrow \infty }\frac{1}{T}
\int_{0}^{T} {\rm Tr}[\rho _{i}(t)H_{i}]\ dt  \label{eq:average}
\end{equation}
where, computationally, we integrate from $0$ up to $20,000/\omega _{s}$
which we have found to be sufficient to ensure the convergence of the
integral (\ref{eq:average}) for all the results presented in this paper. In
figure~4 we display the computed, time averaged, energy expectation values
(normalized in units of $\hbar \omega _{s}$) of $H{{^{\prime }}}_{{s}}$
[figure~4(a)] and $H{{\ ^{\prime }}}_{{e}}$ [figure~4(b)]. These have been
calculated over the range $0\leq \varphi _{x}\leq 1$ for various values of $
\sigma \left( =0,1,2,3\right) $, with $\mu =1/100$ and $\omega _{e}=\omega
_{s}$. In computing these results we have set the $t=0$ state as $|\alpha =i 
\sqrt{2}\rangle _{e}\otimes |\sigma \rangle _{s}$, where $|\alpha \rangle
_{e}$ is a coherent state of the em field ($a_{e}|\alpha \rangle _{e}=\alpha
|\alpha \rangle _{e}$). As is apparent in figures~4(a) and~(b), for specific
values of external bias flux $\varphi _{x}$ [namely those corresponding to
the crossing points shown in figure~3(a)] there is a strong interaction
between the em field and the SQUID ring. As is also apparent, this leads to
an energy exchange between the components of the system. To demonstrate how
the time averaged energy levels for the ring and field depend on the ratio
of $\omega _{s}$ to $\omega _{e}$, we show in figure~5(a) and (b),
respectively, these levels computed for $\sigma =0,1$ and, again, $\mu
=1/100 $ but now $\omega _{e}=\omega _{s}/10$. In order for the energy of
our initial state to be equal to that used in the previous example, here
this state is chosen to be $|\alpha =i10\sqrt{2}\rangle _{e}\otimes |\sigma
\rangle _{s}$. As is to be expected, starting with $\sigma =0,1$
eigenstates, the separation in $\varphi _{x}$ between the regions of strong
coupling (energy exchange) are significantly reduced compared to those seen
in figure~4. As a further example, we show in figure~6 the computed results
for our coupled system taking, as in figure~4, $\omega _{e}=\omega _{s}$ but
now with stronger coupling $\left( \mu =1/10\right) $. To make our results
strictly comparable with those of figure~4, we use the initial state $
|\alpha =i\sqrt{2}\rangle _{e}\otimes |\sigma =0\rangle _{s}$. Due to the
stronger coupling we can see more regions in external bias flux where energy
is exchanged between the two components of the system. In all three sets of
results (figures~4,~5 and~6) there are peaks (both upwards and downwards)
generated in the time averaged energies about specific values of $\varphi
_{x}$. These peak regions, where energy is exchanged between the field and
the ring, correspond to quantum transitions in the ring and in all cases
demonstrate strong coupling between the two oscillators in the system.

\begin{figure}
\protect{  \begin{center}
    \resizebox*{0.48\textwidth}{!}{\includegraphics{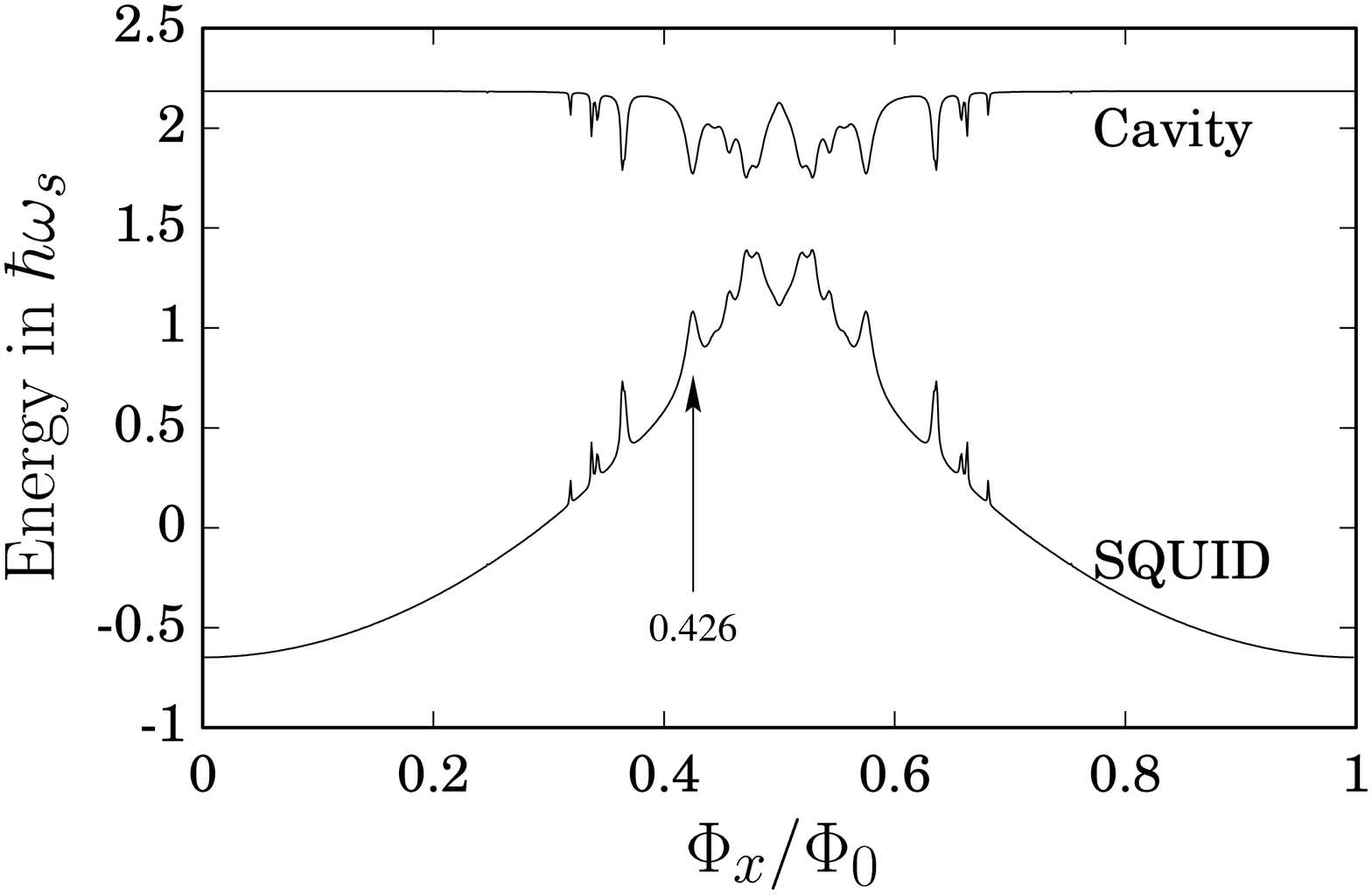}}
  \end{center} 
  \caption{
The time averaged energy levels versus $\varphi _{x}$ for
the ring ($\left\langle H_{s}\right\rangle $) and the field ($\left\langle
H_{e}\right\rangle $). The coupling constant $\mu =1/10$ and $\omega
_{e}=\omega _{s}$. At $t=0$ the electromagnetic field is assumed to be in
the coherent state $|\alpha =i\sqrt{2}\rangle $ and the ring in the lowest
eigenstate $\sigma =0$.
 \label{fig:6}
 }}
\end{figure}

\begin{figure}
\protect{  \begin{center}
    \resizebox*{0.48\textwidth}{!}{\includegraphics{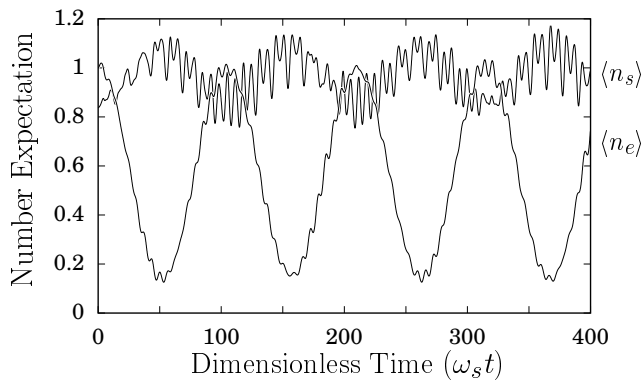}}
  \end{center} 
  \caption{
The expectation values $\left\langle n_{e}\right\rangle $
and $\left\langle n_{s}\right\rangle $ as functions of time for $\varphi
_{x}=0.426,$ $\mu =1/10$ and $\omega _{e}=\omega _{s}$. At $t=0$ the
electromagnetic field is assumed to be in the number state $|1\rangle $ and
the ring is in the energy eigenstate $\sigma =0$.
 \label{fig:7}
 }}
\end{figure}

To illustrate the quantum mechanical effects associated with this coupling
we take as an example the case of $\mu =1/10$, $\omega _{e}=\omega _{s}$ (as
in figure~6) and set $\varphi _{x}=0.426$ (arrowed in figure~6) at which
flux bias the coupling (and the energy exchange) between the ring and field
is strong. We assume that at $t=0$ the em field is in the number state $
|1\rangle $ ($a_{e}^{\dagger }a_{e}|1\rangle _{e}=1|1\rangle _{e}$) and the
ring is in the energy eigenstate $\sigma =0$ (we stress that $\sigma $ are
eigenstates of $H_{s}$ and not of $n_{s}=a_{s}^{\dagger }a_{s}$ ). In
figure~7, with these values of $\omega _{s}$, $\mu $ and $\varphi _{x}$, we
show the computed expectation values of the photon number $\left\langle
n_{e}\right\rangle ={\rm Tr}(\rho _{e}a_{e}^{\dagger }a_{e})$ in the field,
and $\left\langle n_{s}\right\rangle ={\rm \ Tr}(\rho _{s}a_{s}^{\dagger
}a_{s})$ in the ring, as functions of time. These results demonstrate that a
strong exchange in energy takes place quasi-periodically in time between the
ring and the field, i.e. when the photon number expectation value in the
SQUID ring increases that in the em field decreases, and vice versa. We note
that in order to compare these predictions with experiment we would need to
to measure the actual power level of the em field. We also note that with
the ring-field coupling constant known this would allow us to estimate the
em power impinging on the SQUID ring.

As the system evolves in time its two components-oscillator mode and SQUID
ring - become entangled quantum mechanically. In order to quantify this
entanglement we use entropic quantities. For a two mode (field-ring) system $
e-s$ this entanglement can be quantified according to the expression~\cite
{lin73,lieb75,wehrl78,barnett91} 
\begin{equation}
I_{es}=S\left( \rho _{e}\right) +S\left( \rho _{s}\right) -S\left( \rho
\right)  \label{eq:e1}
\end{equation}
where $S\left( \rho \right) $ is the von-Neumann entropy given by 
\begin{equation}
S\left( \rho \right) =-{\rm Tr}\left[ \rho \ln \left( \rho \right) \right]
\end{equation}

\begin{figure}
\protect{  \begin{center}
    \resizebox*{0.48\textwidth}{!}{\includegraphics{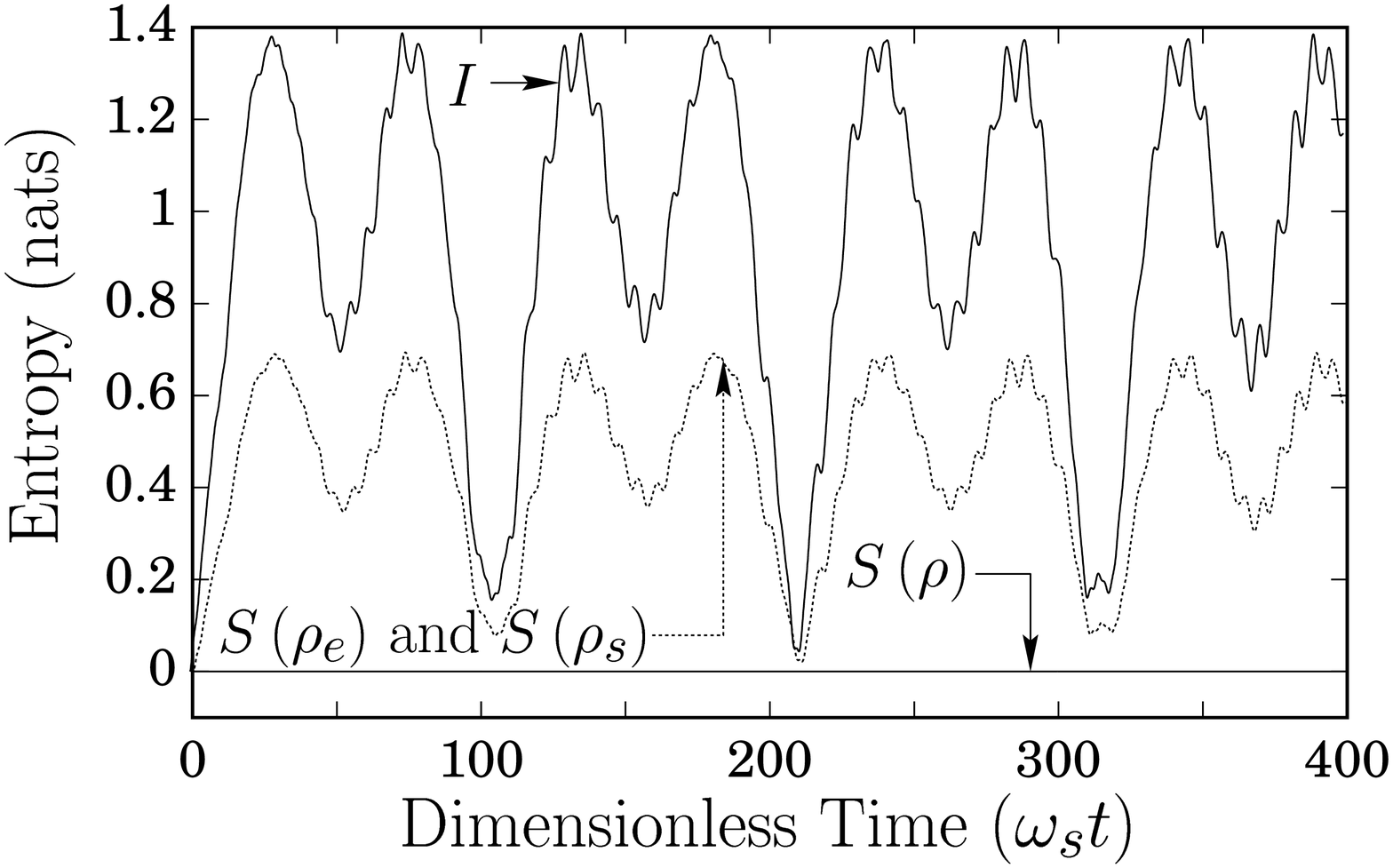}}
  \end{center} 
  \caption{
 The entropies as a function of time for the same system as
in figure 7.
 \label{fig:8}
 }}
\end{figure}
and the entanglement entropy is positive or zero (subadditivity property of
the entropy). In figure~8 we show the time dependent computed entropies $
S\left( \rho _{e}\right) $, $S\left( \rho _{s}\right) $, and the
entanglement entropy $I$, for the same system as in figure~7. From these
results it is quite apparent that although at $t=0$ both the field and the
ring are in a pure state, they both evolve into mixed states. Of course,
since the time evolution is unitary the joint field-ring system is always in
a pure state $\left( S(\rho )=0\right) $. The results presented in figure~8
do demonstrate that the system does become highly entangled over time
although, as can be seen, at certain times it can disentangle again. There
is no doubt that for the development of truly quantum technologies, for
example, quantum computing and quantum communications, such states of
entanglement are of great importance. We note in particular, that in some of
these schemes the ability of the system to control the entanglement (as in
our case) is highly desirable. In principle, experimental verification of
the entanglement between the two modes could be achieved through Bell type
of inequalities. However, in the context of the present work their exact
form will require further investigation.

\section{A SQUID ring coupled to a classical em field}

In previous work we treated the em field classically~\cite{3.} and used the
Hamiltonian 
\begin{equation}
H_{s}=\frac{Q_{s}^{2}}{2C_{s}}+\frac{\left( \Phi _{s}-\left[ \Phi
_{x}+\varphi _{e}\sin (\omega _{e}t)\right] \right) ^{2}}{2\Lambda _{s}}
-\hbar \nu \cos \left( 2\pi \frac{\Phi _{s}}{\Phi _{0}}\right).
\label{eq:HamTDSE}
\end{equation}
and solved the corresponding time-dependent Schr\"{o}dinger equation. Here $
\varphi _{e}$ is taken to be the magnetic flux amplitude of the classical em
field. It is of interest to compare the quasi-classical results derived via
(\ref{eq:HamTDSE}) with the fully quantum results found above where the
initial state of the em field is a coherent state. Due to the
quasi-classical nature of the coherent state we expect some agreement
between the fully quantum results and the quasi-classical results. To
furnish an example to compare with these quantum results, we have computed
the time averaged ring energy expectation values for the Floquet states
(eigenvalues of the evolution operator after one period of microwave
evolution) as a function of $\varphi _{x}$ using the same value of microwave
field amplitude $\left( \varphi _{e}=0.41\mu \Phi _{0}\right) $ as in
figure~4(a). Our results are presented in figure~9. Within the computational
accuracy available, and given that we are dealing with two different regimes
of the coupled system, it is clear that the principal transition region
features match in both models, even though the amplitudes may not be the
same.

\begin{figure}
\protect{  \begin{center}
    \resizebox*{0.48\textwidth}{!}{\includegraphics{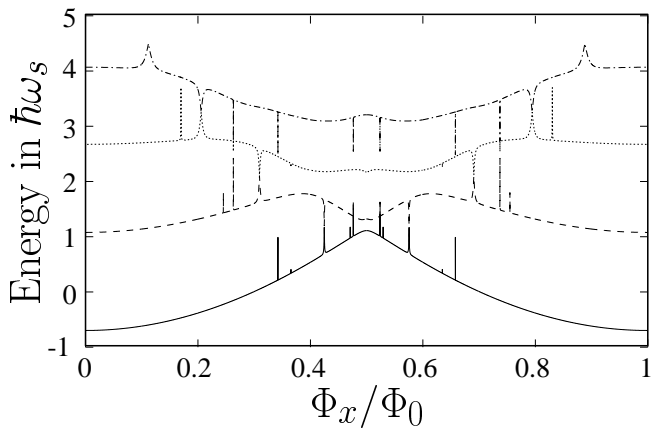}}
  \end{center} 
  \caption{
Time-dependent Schr\"{o}dinger equation (Eq 12) calculation
of the time averaged ring energy levels against $\Phi _{x}$ with $\mu =1/100$
, $\omega _{e}=\omega _{s}$ and $\varphi _{e}=0.41\mu \Phi _{0}$ Here the
electromagnetic field is treated classically and the time-dependent
Schr\"{o}dinger equation (12) is solved. The results of this figure should
be compared and contrasted with those of figure 4a.
 \label{fig:9}
 }}
\end{figure}

\section{Discussion}

We have studied the coupling of a SQUID ring to a single mode em field at
frequencies in the sub-THz to THz range where, in general, $k_{B}T<\hbar
\omega _{s},\hbar \omega _{e}$ so that the system behaves quantum
mechanically. In this we have been strongly influenced by recent discussions
in the literature on routes to quantum computing using Josephson devices\cite
{1.,2.,schon1999,A} and on even more recent publication of experimental data
on superposition states in SQUID rings~\cite{L,M}. In the paper we have
shown that our results [for example, figure~4(a)] compare well with previous
semi-classical (Floquet method) computations made by us (figure~9 and
ref~11) and have expanded this work to calculate explicitly the photon
number and entanglement states in the coupled ring-field system. We note the
relation between the results presented in this paper and previous work on em
environments in thermal equilibrium with Josephson (e.g. SQUID) circuits~
\cite{dev90,girvin90,averin90,flen91,ingold91,falci91,maas91}. In our work
we have assumed that the external em field is in a particular quantum state
and our results depend on this state. As an example we have considered the
em field at $t=0$ to be in a coherent state. However, the calculations can
easily be repeated for another initial state of the field. Each initial
state will, of course, yield different results.

In addition to the em field a magnetostatic flux also threads the SQUID
ring. In this paper we have demonstrated that for certain values of this
flux strong coupling develops at which point(s) large amounts of energy are
exchanged between the ring and the field. Future experimental probing of
these energy exchanges, which is considered again in the discussion section,
would clearly be of great interest. We have also demonstrated that
entanglement between the ring and field modes arises as a natural
consequence of the full quantum mechanics of the system. As we have pointed
out (section II, above), experimental verification of such entanglements
will require further work on the Bell inequalities related to these
entanglements.

In our calculations we have neglected dissipation and have calculated the
time evolution of the system using the equation $\partial _{t}\rho =-\left(
i/\hbar \right) [H,\rho ]$. A more realistic calculation to take into
account dissipation due to the external environment~\cite{Leg} can be made
with the equation $\partial _{t}\rho =-\left( i/\hbar \right) [H,\rho
]+f(\rho )$ where the $f(\rho )$ are dissipative terms. Numerical work to
include these terms is currently in progress.

Of general interest to experimentalists working on (time dependent)
superposition states in SQUID ring devices is the problem of determining the
actual em power (or number of photons in each state of the em field) coupled
to the ring. Together with the frequency of the em field, and the original
eigenenergies of the ring, this is required in order to compute the ring
crossing point splittings. In principle, this problem can be overcome
through the kind of analysis we have undertaken in this paper, whether it be
for classical em fields\cite{3.} or photon states interacting with a quantum
mechanical SQUID ring. As we have shown, it is possible to determine these
power levels accurately through following the reactive frequency shift of a
SQUID ring-classical (radio frequency) resonator system, driven by an
external em field, when the ring remains adiabatically in its quantum
mechanical ground state. However, where em frequencies and/or amplitudes are
high enough (as in this paper), so that (time dependent) superpositions of
low lying energy eigenstates of the ring are generated, the problem becomes
very much more difficult theoretically. Nevertheless, there appear to be
several routes to resolving these difficulties, as indicated by some of our
recent investigations of non-adiabatic processes in em-driven SQUID rings~
\cite{whiteman1997,thesis}. We note that at sufficiently high em
frequencies/amplitudes multiphoton absorption and emission processes will
occur between the components of the coupled system. This may complicate the
interpretation of experimental data and will be a topic of further
theoretical investigation by us. For the future, we also note that it may be
possible to extend these experimental and theoretical techniques to probe
the details of energy exchange and entanglement of the system presented in
this paper.

There now exists a clearly defined need to create THz technology~\cite{pep00}
for a range of applications including modern communications. To date this
technology, based on quantum processes on the small scale, functions
classically at the device level. A primary purpose of this paper has been to
demonstrate that at THz frequencies, and reasonable operating temperatures $
\left( \symbol{126}4K\right) $, this technology could be made fully quantum
mechanical in nature, i.e. at high enough SQUID ring and em oscillator
frequencies both can be treated as macroscopic quantum objects, irrespective
of any deeper description of the superconducting condensate in SQUID rings~
\cite{ralph1996,ralph1997}. This would point to a great richness of
potential applications. For example, in the context of the results presented
here, our investigations may prove useful for the development of frequency
converters up to THz frequencies and beyond. More generally these results,
and the theoretical descriptions underlying them, may find use in the
emerging fields of quantum technology and quantum computation~\cite
{1.,2.,schon1999,A}.

\begin{acknowledgments}

We would like to thank NESTA for its generous funding of this work. We would
also like to express our gratitude to Professors C.H. van de Wal and
A.Sobolev for interesting and informative discussions.
\end{acknowledgments}

\end{document}